\documentclass{iau_FM}
\usepackage{graphicx}

\graphicspath{{.}{./FIGS/}{..}}

\title[Understanding mechanical feedback from HERGs and LERGs] 
{Understanding mechanical feedback from HERGs and LERGs} 

\author[Imogen H. Whittam]   
{Imogen H. Whittam$^1$
 }

\affiliation{$^1$Department of Physics and Astronomy, University of the Western Cape, \\
Robert Sobukwe Road, Bellville 7535, South Africa\\email: {\tt iwhittam@uwc.ac.za}}

\pubyear{2018}
\jname{Astronomy in Focus, Volume 1} 
\editors{Volker Beckmann \& Claudio Ricci, eds.}
\begin{document}
\maketitle

\begin{abstract}
The properties of $\sim 1000$ high-excitation and low-excitation radio galaxies (HERGs and LERGs) selected from the \cite{2016MNRAS.460.4433H} $1 - 2$~GHz VLA survey of Stripe 82 are investigated. The HERGs in this sample are generally found in host galaxies with younger stellar populations than LERGs, consistent with other work. The HERGs tend to accrete at a faster rate than the LERGs, but there is more overlap in the accretion rates of the two classes than has been found previously. We find evidence that mechanical feedback may be significantly underestimated in hydrodynamical simulations of galaxy evolution; 84 \% of this sample release more than 10 \% of their energy in mechanical form. Mechanical feedback is significant for many of the HERGs in this sample as well as the LERGs; nearly 50 \% of the HERGs release more than 10 \% of their energy in their radio jets.
\keywords{radio continuum: galaxies, galaxies: active, galaxies: evolution}
\end{abstract}

\firstsection 
\section{Introduction}

One of the key unknowns in galaxy evolution is how star-formation in galaxies becomes quenched; it is widely thought that feedback from active galactic nuclei (AGN) is responsible for this, but the mechanisms are not well understood. Observational evidence (e.g. \cite{2012MNRAS.421.1569B}) suggests that AGN can be split into two distinct classes; high-excitation radio galaxies (HERGs; also known as cold mode, quasar mode or radiative mode sources) which radiate efficiently across the electromagnetic spectrum and posses the typical AGN accretion-related structures such as an accretion disk and a dusty torus, and low-excitation radio galaxies (LERGs; also known as hot mode, radio mode or jet mode sources) which radiate inefficiently and emit the bulk of their energy in mechanical form as powerful radio jets (see e.g.\ \cite{2014ARA&A..52..589H}). 

It is thought that these two AGN accretion modes have different feedback effects on the host galaxy (see review by \cite{2012ARA&A..50..455F}) and lead to the two different feedback paths in semi-analytic and hydrodynamic simulations, but despite being widely studied over the last decade (e.g. \cite{2007MNRAS.376.1849H}; \cite{2009Natur.460..213C}; \cite{2014ARA&A..52..589H}) these processes are not well understood. 
In these proceedings I use a sample of $\sim 1000$ HERGs and LERGs to investigate the host galaxy properties and accretion rates of the two classes, and explore the implications of these results for AGN feedback.

\section{Data used and source classification}

This work is based on a $1-2$~GHz Karl G. Jansky Very Large Array (VLA) survey covering 100~deg$^2$ in SDSS Stripe 82 which has a $1 \sigma$ rms noise of 88~$\mu$Jy beam$^{-1}$ and a resolution of $16 \times 10$~arcsec; full details of the radio survey are given in \cite{2016MNRAS.460.4433H}.  This radio catalogue was matched to the SDSS DR14 optical catalogue (\cite{2018ApJS..235...42A}) by eye; details of the matching process are described in \cite{2018MNRAS.480..707P}. We restrict our analysis to sources with a counterpart in the spectroscopic catalogue with $z < 0.7$, our sample therefore has 1501 sources which cover the range $0.01 < z < 0.7$ and $10^{21} < L_{1.4~\rm GHz} / \textrm{W Hz}^{-1} < 10^{27}$. We use the the value-added spectroscopic catalogues described in \cite{2013MNRAS.431.1383T}. 

Sources are classified as either AGN or star-forming galaxies using information from their optical spectra, full details of this process are given in \cite{2018MNRAS.480..707P}. The AGN in the sample are then classified as HERGs or LERGs using the criteria given in \cite{2012MNRAS.421.1569B}, which use a combination of emission line ratios and [OIII] equivalent width. This is explained in detail in \cite{2018MNRAS.480..358W}. Additionally to the Best and Heckman classification scheme, we introduce a `probable LERG' class for sources which cannot be classified according to the full criteria but which have an [OIII] equivalent width $<5$~\AA. 
The total number of sources in each category is as follows; HERGs = 60, LERGs = 149, probable LERGs = 600, QSOs = 81 and unclassified sources = 271, with 340 star-forming galaxies.

\section{Host galaxy properties}

\begin{figure}
\centerline{\includegraphics[width=7cm]{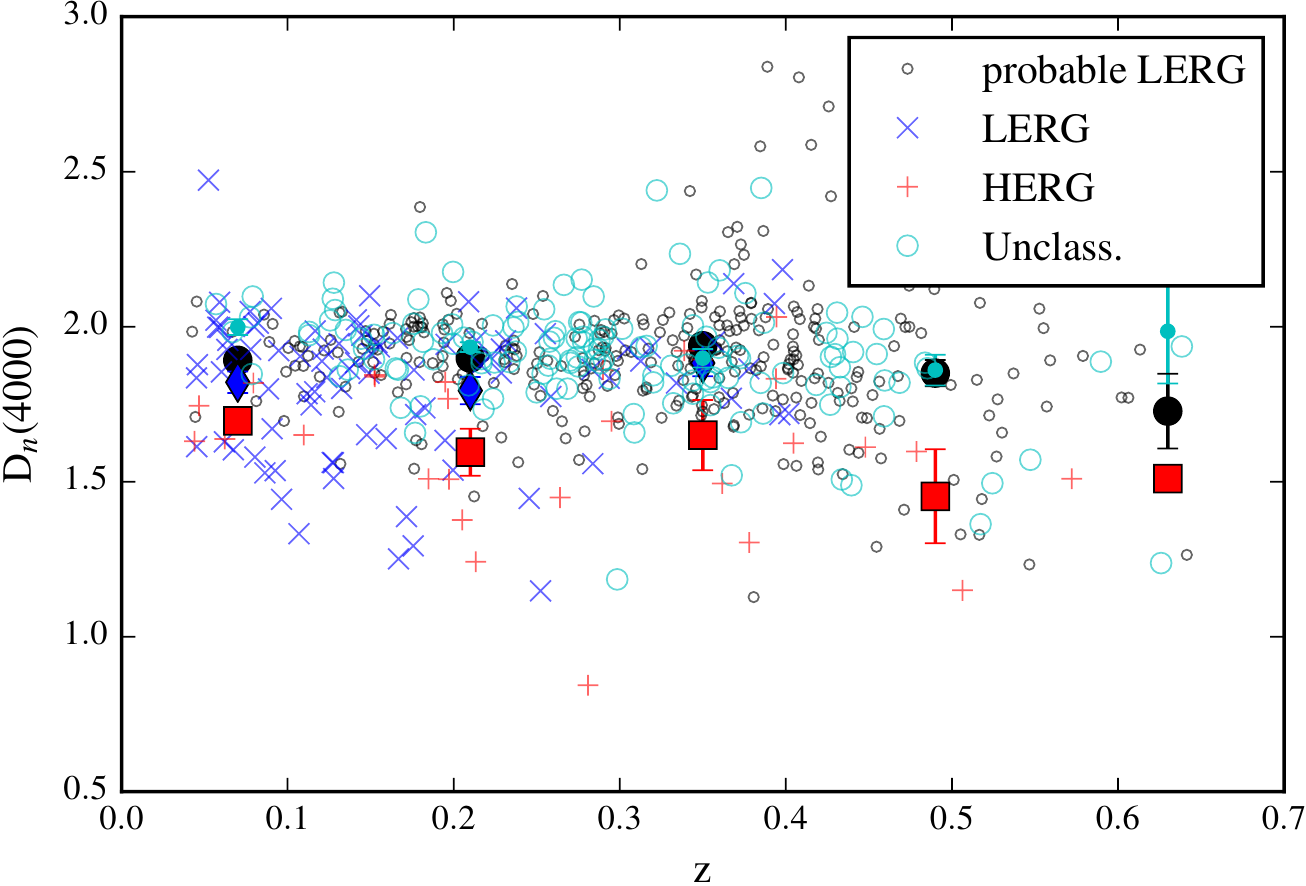}}
\caption{4000~\AA~break strength as a function of redshift with the HERGs, LERGs, probable LERGs and unclassified sources shown separately. The filled shapes show the mean values in each luminosity bin for the different samples. From \cite{2018MNRAS.480..358W}}\label{fig:Dn4000}
\end{figure}

Using the wealth of multi-wavelength data available in the field, we can compare the properties of the host galaxies of the HERGs and LERGs. 4000~\AA~break strength, which traces stellar age, is shown as a function of redshift in Fig~\ref{fig:Dn4000}. This shows that HERGs tend to be found in host galaxies with younger stellar populations than LERGs across the redshift range probed here. This is agrees with other results in the literature (e.g.\ \cite{2012MNRAS.421.1569B}) and is consistent with the idea that HERGs have a supply of cold gas which provides the fuel for both star-formation and AGN activity. We refer the reader to \cite{2018MNRAS.480..358W} for further discussion of this and other host galaxy properties.

\section{Accretion rates}

\begin{figure}
\centerline{\includegraphics[width=6.5cm]{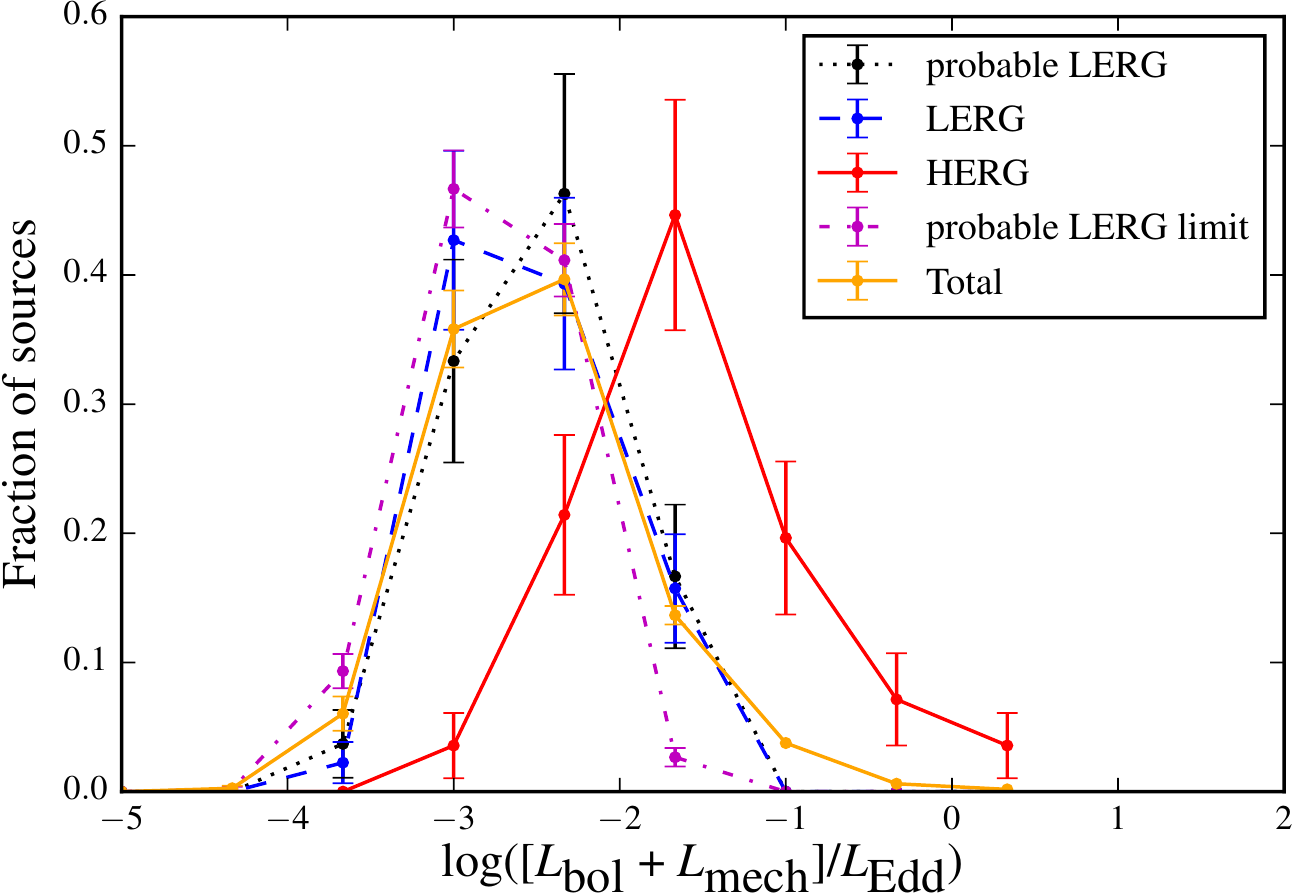}
            \quad
            \includegraphics[width=6.5cm]{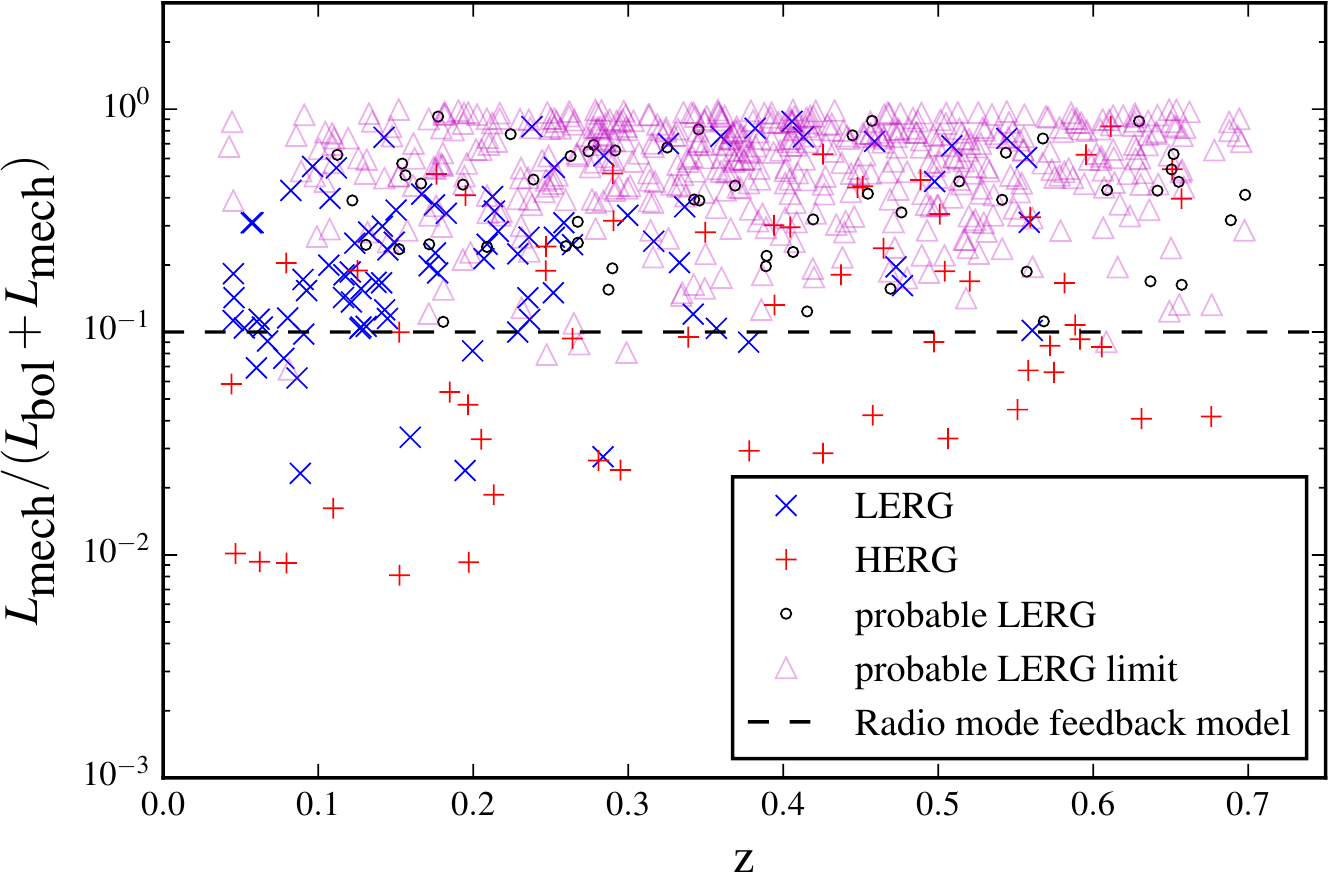}}
\caption{Left panel shows the distribution of Eddington-scaled accretion rates for the different source classifications.  Right panel shows the fraction of the accreted energy released in the jets for the different source types as a function of redshift. Triangles represent sources with an upper limit on their radiative accretion rate, so the fraction of energy released in the jet is a lower limit. The dashed line is the radio mode feedback model used in Horizon-AGN from \cite{2014MNRAS.444.1453D}. The uncertainties in the scaling relations used to estimate $L_\textrm{bol}$ and $L_\textrm{mech}$ are 0.4 and 0.7 dex respectively. From \cite{2018MNRAS.480..358W}. }\label{fig:accretion}
\end{figure}

There is a scenario building up in the literature that there are two distinct accretion modes which are responsible for HERGs and LERGs respectively; in this scenario there is a dichotomy in accretion rates between the two classes, relating to the two different modes.
The radiative accretion rates of the AGN in this sample are estimated from their [OIII] 5007 line luminosity and the mechanical accretion rates are estimated from the 1.4-GHz radio luminosity using the \cite{2010ApJ...720.1066C} relationship. Black hole masses are estimated from the local black hole mass - bulge mass relation, allowing Eddington-scaled accretion rates to be calculated as follows: $\lambda = (L_{\rm bol} + L_{\rm mech}) / L_{\rm Edd}$. The left panel of Fig.~\ref{fig:accretion} shows the distribution of Eddington-scaled accretion rates for the HERGs and LERGs in this sample.  It is clear from this figure that the HERGs generally accrete at a faster Eddington-scaled rate than the LERGs, with a distribution that peaks just below 0.1 compared to 0.01. However, there is a significant overlap in accretion rates between the two classes, with HERGs found across nearly the full range of accretion rates.

The dichotomy in accretion rates between HERGs and LERGs is therefore less clear in this study than it is in other studies in the literature; for example \cite{2012MNRAS.421.1569B} and \cite{2014MNRAS.440..269M} both find almost no overlap in accretion rates between the two classes. In contrast, our sample seems to suggest a more continuous range of accretion rates. Note the our sample probes fainter radio luminosities ($10^{21} < L_{1.4~\rm GHz} / \textrm{W Hz}^{-1} < 10^{27}$) than other results in the literature; this could be part of the reason for the difference in our results, although we see some overlap in the accretion rates of the HERGs and LERGs across the luminosity range sampled here. We also do not observe any dichotomy in the [OIII] equivalent width or Excitation Index distributions, the two main parameters used to classify the HERGs and LERGs, suggesting that any dividing value chosen in these parameters is perhaps arbitrary for our sample.

\section{Implications for AGN feedback}

AGN feedback is required in all leading hydrodynamical simulations of galaxy evolution to quench star-formation. Some simulations implement mechanical and radiative feedback (assumed to relate to LERGs and HERGs respectively) separately (e.g. Horizon-AGN; \cite{2014MNRAS.444.1453D}) while others do not (e.g. MUFASA; \cite{2016MNRAS.462.3265D}).

The right panel of Fig.~\ref{fig:accretion} shows $L_\textrm{mech} / (L_\textrm{bol} + L_\textrm{mech})$, which provides an estimate of the fraction of the total accreted energy deposited back into the interstellar medium in mechanical form. The dashed line shows the mechanical feedback efficiency of 10 \% assumed in Horizon-AGN; it is clear that this is a significant underestimate for the sources in this sample, with 84 \% of the sample depositing more than 10 \% of their energy in mechanical form.
This plot also demonstrates that mechanical feedback can be significant for HERGs as well as for LERGs; nearly 50 \% (29/60) of the HERGs in this sample release more than 10 \% of their accreted energy in mechanical form. 

There is a scatter of $\sim 2$~dex in $L_\textrm{mech} / (L_\textrm{bol} + L_\textrm{mech})$, which shows that the assumption that there is a direct scaling between accretion rate and mechanical feedback which is used in most hydrodynamical simulations does not necessarily hold. This may be because environment plays a significant role.

\section{Conclusions and future perspectives}

We have used the \cite{2016MNRAS.460.4433H} VLA 1-2 GHz radio survey covering 100~deg$^2$ in Stripe 82 along with optical spectroscopy to probe the properties of $\sim 1000$ high- and low-excitation radio galaxies. They key results of this work are:
\begin{itemize}
  \item HERGs tend to be found in host galaxies with younger stellar populations than LERGs, consistent with other results in the literature.
  \item While the HERGs in our sample tend to have higher accretion rates than the LERGs, we find considerable overlap in the accretion rates of the two samples.
  \item Mechanical feedback can be significant for HERGs as well as for LERGs, and may be underestimated for both populations in hydrodynamical simulations.
\end{itemize}

The advent of new radio telescopes, such as MeerKAT, LOFAR and ASKAP, means there is potential to make a large step forward in our understanding of radio galaxies and their mechanical feedback effects in the next few years. One example of a survey planned with a new instrument is the MeerKAT MIGHTEE survey (\cite{2016mks..confE...6J}) which has just started to collect data and will survey 10~deg$^2$ to a depth of 1~$\mu$Jy at 800 - 1600 MHz in four different fields. The unique combination of deep radio images over a significant cosmological volume along with excellent multi-wavelength coverage means we will be able to, amongst other things, extend the study described in this proceedings to significantly fainter luminosities and probe whether or not there is an accretion mode dichotomy, particularly at lower luminosities.

\smallskip

\noindent\textit{Acknowledgements} The author thanks Matthew Prescott, Matt Jarvis, Kim McAlpine and Ian Heywood for their significant contributions to this work. This research was supported by the South African Radio Astronomy Observatory, which is a facility of the National Research Foundation, an agency of the Department of Science and Technology.


\begin{thebibliography}{}

\bibitem[Abolfathi et al.\ (2018)]{2018ApJS..235...42A} 
  Abolfathi B., et al., 2018, \textit{ApJS}, 235, 42 

 \bibitem[Best \& Heckman (2012)]{2012MNRAS.421.1569B} 
  Best P.~N., Heckman T.~M., 2012, \textit{MNRAS}, 421, 1569 

 \bibitem[Cattaneo et al.\ (2009)]{2009Natur.460..213C} 
  Cattaneo A., et al., 2009, \textit{Nature}, 460, 213 

 \bibitem[Cavagnolo et al.\ (2010)]{2010ApJ...720.1066C} 
  Cavagnolo K.~W., et al., 2010, \textit{ApJ}, 720, 1066 

 \bibitem[Dav{\'e} et al.\ (2016)]{2016MNRAS.462.3265D} 
  Dav{\'e} R., Thompson R., Hopkins P.~F., 2016, \textit{MNRAS}, 462, 3265 

 \bibitem[Dubois et al.\ (2014)]{2014MNRAS.444.1453D} 
  Dubois Y., et al., 2014, \textit{MNRAS}, 444, 1453 

 \bibitem[Fabian (2012)]{2012ARA&A..50..455F} 
  Fabian A.~C., 2012, \textit{ARA\&A}, 50, 455 

 \bibitem[Hardcastle et al.\ (2007)]{2007MNRAS.376.1849H} 
  Hardcastle M.~J., Evans D.~A., Croston J.~H., 2007, \textit{MNRAS}, 376, 1849 

 \bibitem[Heckman \& Best (2014)]{2014ARA&A..52..589H} 
  Heckman T.~M., Best P.~N., 2014, \textit{ARA\&A}, 52, 589 

 \bibitem[Heywood et al.\ (2016)]{2016MNRAS.460.4433H} 
  Heywood I., et al., 2016, \textit{MNRAS}, 460, 4433 

 \bibitem[Jarvis et al.\ (2016)]{2016mks..confE...6J} 
  Jarvis M., et al., 2016, \textit{Proceedings of MeerKAT Science: On the Pathway to the SKA. 25-27 May, 2016 Stellenbosch, South Africa}, 6 

 \bibitem[Mingo et al.\ (2014)]{2014MNRAS.440..269M} 
  Mingo B., et al., 2014, \textit{MNRAS}, 440, 269 

 \bibitem[Prescott et al.\ (2018)]{2018MNRAS.480..707P} 
  Prescott M., et al., 2018, \textit{MNRAS}, 480, 707 

 \bibitem[Thomas et al.\ (2013)]{2013MNRAS.431.1383T} 
  Thomas D., et al., 2013, \textit{MNRAS}, 431, 1383 

 \bibitem[Whittam et al.\ (2018)]{2018MNRAS.480..358W} 
  Whittam I.~H., Prescott M., McAlpine K., Jarvis M.~J., Heywood I., 2018, \textit{MNRAS}, 480, 358 


\end{thebibliography}
\end{document}